\documentclass[sigconf]{acmart}
\AtBeginDocument{%
  \providecommand\BibTeX{{%
    \normalfont B\kern-0.5em{\scshape i\kern-0.25em b}\kern-0.8em\TeX}}}

\setcopyright{rightsretained}
\copyrightyear{2023}
\acmYear{2023}
\setcopyright{rightsretained}
\acmConference[CIKM '23]{Proceedings of the 32nd ACM International
Conference on Information and Knowledge Management}{October 21--25,
2023}{Birmingham, United Kingdom}
\acmBooktitle{Proceedings of the 32nd ACM International Conference on
Information and Knowledge Management (CIKM '23), October 21--25, 2023,
Birmingham, United Kingdom}\acmDOI{10.1145/3583780.3614844}
\acmISBN{979-8-4007-0124-5/23/10}

\makeatletter
\gdef\@copyrightpermission{
  \begin{minipage}{0.3\columnwidth}
   \href{https://creativecommons.org/licenses/by/4.0/}{\includegraphics[width=0.9\textwidth]{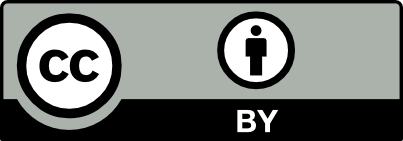}}
  \end{minipage}\hfill
  \begin{minipage}{0.7\columnwidth}
   \href{https://creativecommons.org/licenses/by/4.0/}{This work is licensed under a Creative Commons Attribution International 4.0 License.}
  \end{minipage}
  \vspace{5pt}
}
\makeatother

\usepackage{enumitem}
\usepackage{array,multirow,graphicx}
\usepackage{float}
\usepackage{xcolor,colortbl}

\newcommand{\ie}{\emph{i.e., }}
\newcommand{\eg}{\emph{e.g., }}

\newcommand{\etc}{\emph{etc. }}

\begin{document}

\title{Diffusion Variational Autoencoder for Tackling Stochasticity in Multi-Step Regression Stock Price Prediction}

\author{Kelvin J.L. Koa}
\affiliation{%
  \institution{National University of Singapore}
  \country{}
  }
\email{kelvin.koa@u.nus.edu}

\author{Yunshan Ma}
\authornote{Corresponding author.}
\affiliation{%
  \institution{National University of Singapore}
  \country{}
  }
\email{yunshan.ma@u.nus.edu}

\author{Ritchie Ng}
\affiliation{%
  \institution{Eastspring Investments, Singapore}
  \country{}
  }
\email{ritchie.ng@eastspring.com}

\author{Tat-Seng Chua}
\affiliation{%
  \institution{National University of Singapore}
  \country{}
  }
\email{dcscts@nus.edu.sg}


\renewcommand{\shortauthors}{Kelvin J.L. Koa, Yunshan Ma, Ritchie Ng, \& Tat-Seng Chua}

\begin{abstract}
Multi-step stock price prediction over a long-term horizon is crucial for forecasting its volatility, allowing financial institutions to price and hedge derivatives, and banks to quantify the risk in their trading books. Additionally, most financial regulators also require a liquidity horizon of several days for institutional investors to exit their risky assets, in order to not materially affect market prices. However, the task of multi-step stock price prediction is challenging, given the highly stochastic nature of stock data. Current solutions to tackle this problem are mostly designed for single-step, classification-based predictions, and are limited to low representation expressiveness. The problem also gets progressively harder with the introduction of the target price sequence, which also contains stochastic noise and reduces generalizability at test-time.

To tackle these issues, we combine a deep hierarchical variational-autoencoder (VAE) and diffusion probabilistic techniques to do seq2seq stock prediction through a stochastic generative process. The hierarchical VAE allows us to learn the complex and low-level latent variables for stock prediction, while the diffusion probabilistic model trains the predictor to handle stock price stochasticity by progressively adding random noise to the stock data. To deal with the additional stochasticity in the target price sequence, we also augment the target series with noise via a coupled diffusion process. We then perform a denoising process to "clean" the prediction outputs that were trained on the stochastic target sequence data, which increases the generalizability of the model at test-time. Our Diffusion-VAE (D-Va) model is shown to outperform state-of-the-art solutions in terms of its prediction accuracy and variance. Through an ablation study, we also show how each of the components introduced helps to improve overall prediction accuracy by reducing the data noise. Most importantly, the multi-step outputs can also allow us to form a stock portfolio over the prediction length. We demonstrate the effectiveness of our model outputs in the portfolio investment task through the Sharpe ratio metric and highlight the importance of dealing with different types of prediction uncertainties. Our code can be accessed through https://github.com/koa-fin/dva.
\end{abstract}



\maketitle

\vspace{-5px}
\section{Introduction}
The stock market, an avenue in which investors can purchase and sell shares of publicly-traded companies, have seen a total market capitalization of over \$111 trillion in 2022\footnotemark{}\footnotetext{https://www.sifma.org/resources/research/research-quarterly-equities/}. Accurate stock price forecasting help investors to make informed investment decisions, allow financial institutions to price derivatives, and let regulators manage the amount of risk in the financial system. As such, the task of stock market prediction has become an increasingly challenging and important one in the field of finance, which has attracted significant attention from both the academia and industry \cite{feng2019temporal, lin2021learning, yang2022numhtml}. 

\vspace{-8px}
\begin{figure}[h!]
\includegraphics[width=\linewidth]{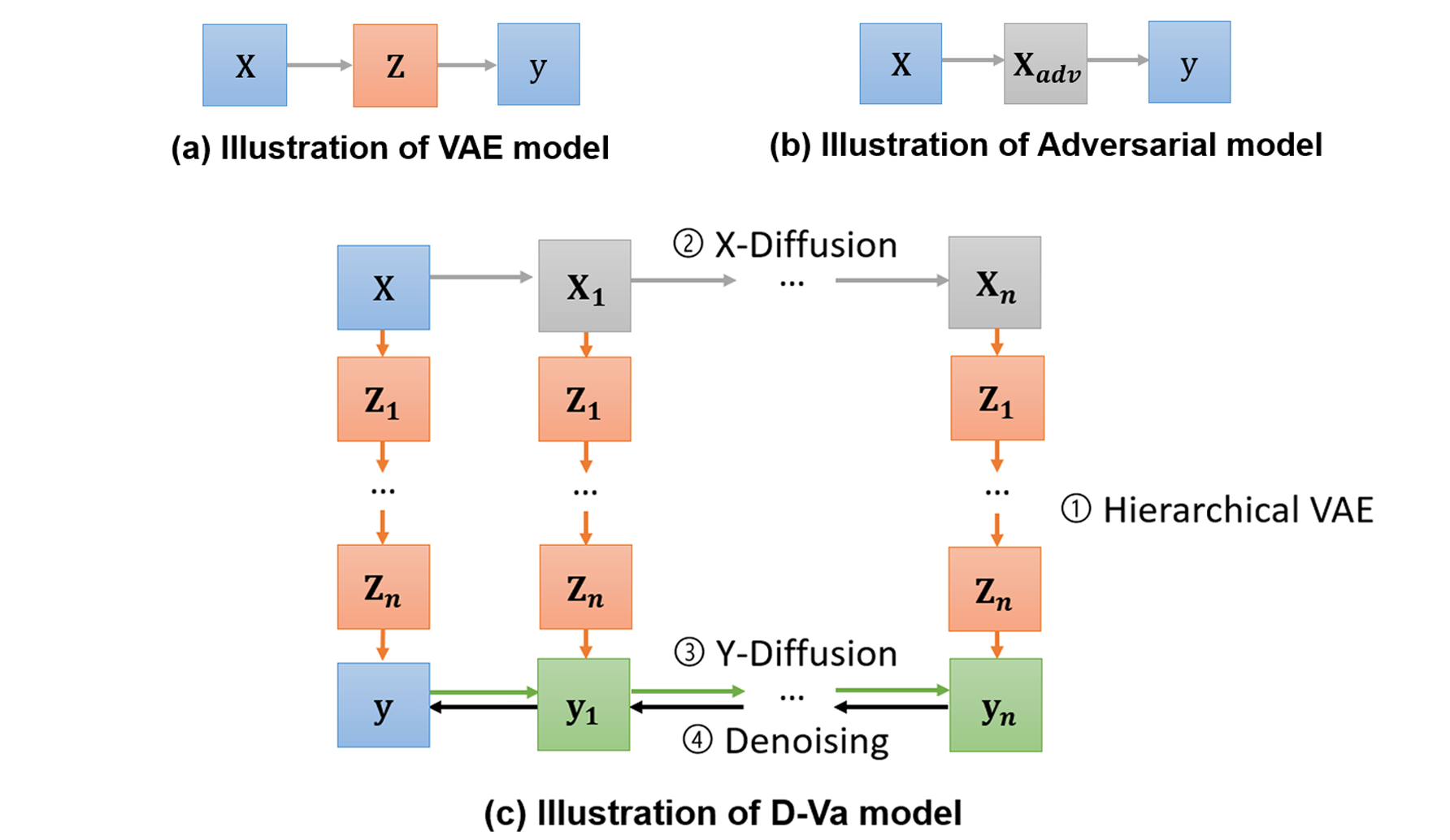}
\vspace{-20px}
\caption{Illustration of the StockNet (VAE) model \cite{xu2018stock}, the Adversarial-ALSTM (Adversarial) model \cite{feng2018enhancing}, and our proposed D-Va model. For the VAE and Adversarial models, $y$ refers to a single-step binary target, while for the D-Va model, $\mathbf{y}$ refers to a multi-step regression sequence target.}
\label{existing_models}
\vspace{-7px}
\end{figure}

Typically, most current works on stock prediction do single-day prediction \cite{hu2018listening, chen2018incorporating}, instead of making predictions for the next multiple days. Intuitively, this allows them to make trading decisions on whether to buy or sell a stock to achieve profits for the next day. However, stock prediction over a longer-term horizon is also crucial to forecast its volatility, which allows for applications such as pricing and hedging financial derivatives by financial institutions and quantifying the risk in banks’ trading books \cite{raunig2006longer}. Additionally, most financial regulators require a liquidity horizon of minimum 10 days for institutional investors to exit their risky assets, in order to not materially affect market prices\footnotemark{}\footnotetext{
Most financial regulatory authorities set a regulatory liquidity horizon chosen from \{10, 20, 40, 60, 120\} days for different asset classes. Some references can be found at:
\begin{itemize}[leftmargin=*]
\item https://www.mas.gov.sg/publications/consultations/2021/consultation-paper-on-draft-standards-for-market-risk-capital-and-capital-reporting-requirements
\item https://www.eba.europa.eu/regulation-and-policy/market-risk/draft-technical-standards-on-the-ima-under-the-frtb
\end{itemize}
}. Currently, there is a limited number of works that do multi-step stock price prediction \cite{dong2013one, liu2018numerical}. Our work fills this gap by designing a method to tackle this task.
 
For our multi-step stock prediction task, two main challenges are identified. Firstly, it is well-established in previous stock prediction literature that stock prices are highly stochastic and standard prediction models tend not to generalize well when dealing with such data \cite{xu2018stock, feng2018enhancing}. Given that stock prices are continuous and change at high frequencies, the discrete stock price data used for training models are essentially stochastic “samples” drawn at specific time-steps, \eg at 12.00am on each day or at the 60-second mark of each minute, which might not fully capture the intrinsic stock price behavior. Currently, existing techniques that tackle this problem include (see Figure \ref{existing_models}): learning a continuous, “latent” variable using a VAE model \cite{kingma2013auto} as the new independent variable for predicting the stock movement \cite{xu2018stock}; and adding adversarial perturbations \cite{goodfellow2020generative} to the stock training data to simulate stock price stochasticity \cite{feng2018enhancing}. However, these models are designed for single-step classification-based stock movement prediction, and are limited to low representation expressiveness. Secondly, for the multi-step regression task, the problem becomes progressively harder with the introduction of the target price sequence. The target series also contains stochastic noise, and training the model to predict this series directly would also reduce the generalizability of the predictions at test-time. 

To deal with the abovementioned problems, we propose our Diffusion-VAE (D-Va) model, which combines deep hierarchical VAEs \cite{sonderby2016ladder, klushyn2019learning, vahdat2020nvae} and diffusion probabilistic \cite{sohl2015deep, song2019generative, ho2020denoising} techniques to do seq2seq stock prediction (see Figure \ref{existing_models}). Firstly, the deep hierarchical VAE increases the expressiveness of the approximate posterior stock price distribution, allowing us to learn more complex and low-level latent variables. Simultaneously, the diffusion probabilistic model trains the predictor to handle stock price stochasticity by progressively adding random noise to the input stock data (X-Diffusion in Figure \ref{existing_models}). Secondly, we deal with the stochastic target price sequence by additionally augmenting the target series with noise via a coupled diffusion process (Y-Diffusion in Figure \ref{existing_models}). The predicted diffused targets are then "cleaned" with a denoising process \cite{da2019style, ma2022denoised} to obtain the generalized, "true" target sequence. This is done by performing denoising score-matching on the diffused target predictions with the actual targets during training, then applying a single-step gradient denoising jump at test-time \cite{saremi2019neural, li2019learning, jolicoeur2020adversarial}. This process can also be seen as removing the estimated aleatoric uncertainty resulting from the data stochasticity \cite{li2023generative}.

To demonstrate the effectiveness of D-Va, we perform extensive experiments, and show that our model is able to outperform state-of-the-art models in overall prediction accuracy and variance. Additionally, we also test the model in a practical investment setting. Having predicted \textit{sequences} of stock returns allow us to form a stock portfolio across the duration of the prediction length using their means and covariance information. Using standard Markowitz mean-variance optimization, we then calculate the portfolio weights of each stock that can maximize the overall expected returns of the portfolio and minimizes its volatility \cite{selection1952harry}. We further regularize the prediction covariance matrix via the graphical lasso \cite{friedman2008sparse} in order to reduce the impact of the uncertainty of the prediction model, which can be seen as a form of epistemic uncertainty \cite{hullermeier2021aleatoric}. We show that tackling both the uncertainties of the data (aleatoric) and the model (epistemic) allow us to achieve the best portfolio performance, in terms of the Sharpe ratio \cite{sharpe1998sharpe}, over the specified test period.

The main contributions of this paper are summarized as:
\begin{itemize}[leftmargin=*]
\item We investigate the problem of generalization in the stock prediction task under the multi-step regression setting, and deal with stochasticity in both the input and target sequences. 

\item We propose a solution that integrates a hierarchical VAE model, a stochastic diffusion process and a denoising component and implement it in an end-to-end model for stock prediction.

\item We conduct extensive experiments on public stock price data across three different time periods, and show that D-Va provide improvements over state-of-the-art methods in prediction accuracy and variance. We further demonstrate the effectiveness of the model in a practical, stock investment setting. 
\end{itemize}
\vspace{-5px}

\section{Related Works}
The task of stock prediction is popular and there is a large amount of literature on this topic. Among the literature, we can classify them into various specific categories in order to position our work:

\textit{Technical and Fundamental Analysis.}
Technical Analysis (TA) methods focus on predicting the future movement of stock prices from quantitative market data, such as the price and volume. Common techniques include using attention-based Long Short-Term Memory (LSTM) networks \cite{qin2017dual}, Autoregressive models \cite{li2016stock} or Fourier Decomposition \cite{zhang2017stock}. On the other hand, Fundamental Analysis (FA) seeks information from external data sources to predict price, such as news \cite{hu2018listening}, earnings calls \cite{yang2022numhtml} or relational knowledge graphs \cite{feng2019temporal}. For our work, we focus on TA to evaluate our techniques on processing quantitative financial data. 

\textit{Classification and Regression.} 
The stock prediction task can be formulated as a binary classification task, where the goal is to predict if prices will move \textit{up} or \textit{down} in the next time-step \cite{ding2020hierarchical}. This is generally considered a more achievable task \cite{ye2021multi} and is sufficient to help retail investors decide whether to buy or sell a stock. On the other hand, one can also formulate it as a regression task and predict the stock price directly. This offers investors more information for decision-making, such as being able to rank stocks based on profits and buying the top percentile stocks \cite{feng2019temporal, lin2021learning}. In this work, we tackle the regression task, in order to be able to \textit{weigh} the amount of each stock for an investment portfolio.

\textit{Single and Multiple Steps Prediction.}
Most current works on stock prediction do single-step prediction for the next time-step \cite{tuncer2022asset}, given that it allows one to make immediate trading decisions for the next day. On the other hand, there are very few literature on multi-step prediction, where stock predictions are made for the next multiple time-steps. An example can be found in \cite{liu2018numerical}, where the authors perform multi-step prediction to analyze the impact of breaking news on stock prices over a period of time. For this work, we will deal with the multi-step prediction task, where the motivation is to allow larger institutions to make long-term, volatility-aware investment decisions. Note that this is different from doing a single time-step prediction over a longer-term horizon \cite{feng2021time}, which can be considered as a single-step prediction task.

\section{Methodology}
In this section, we first formulate the task for the multi-step regression stock prediction. We then present the proposed D-Va model, which is illustrated in Figure \ref{model}. There are four main components in the framework: (1) a hierarchical VAE to generate the sequence predictions; (2) a diffusion process that gradually adds Gaussian noise to the input sequence to simulate stock price stochasticity; (3) a coupled diffusion process additionally applied to the target sequence; and (4) a denoising function that is trained to “clean” the predictions by removing the stochasticity from the predicted series. We will elaborate on each of these components in more details.

\vspace{-5px}
\subsection{Problem Formulation}
For each stock s, given an input sequence of $T$ trading days $\mathbf{X} = \{\mathbf{x}_{t-T+1}, \mathbf{x}_{t-T+2}, \cdots,  \mathbf{x}_{t}\}$, we aim to predict its returns sequence over the next $T'$ trading days $\mathbf{y} = \{r_{t+1}, r_{t+2}, \cdots,  r_{t+T'}\}$, where $r_{t}$ refers to its percentage returns at time $t$, i.e. $r_{t} = c_{t} / c_{t-1}$, and $c$ is the closing price. The input vector consist of its open price, high price, low price, volume traded, absolute returns and percentage returns, i.e. $\mathbf{x}_{t} = \left [o_{t}, h_{t}, l_{t}, v_{t}, \Delta_{t}, r_{t}\right ]$, making it a multi-variate input to single-variate output prediction task. Additionally, given that we are interested in predicting the percentage returns, the open, high and low prices are normalized by the previous closing price, \eg $o_{t} = o_{t} / c_{t-1}$, similar to what was done in \cite{xu2018stock, feng2018enhancing}. The absolute returns $\Delta_{t} = c_{t} - c_{t-1}$ is also included as an input feature.

\vspace{-5px}
\subsection{Deep Hierarchical VAE}
\begin{figure*}
\includegraphics[width=\textwidth]{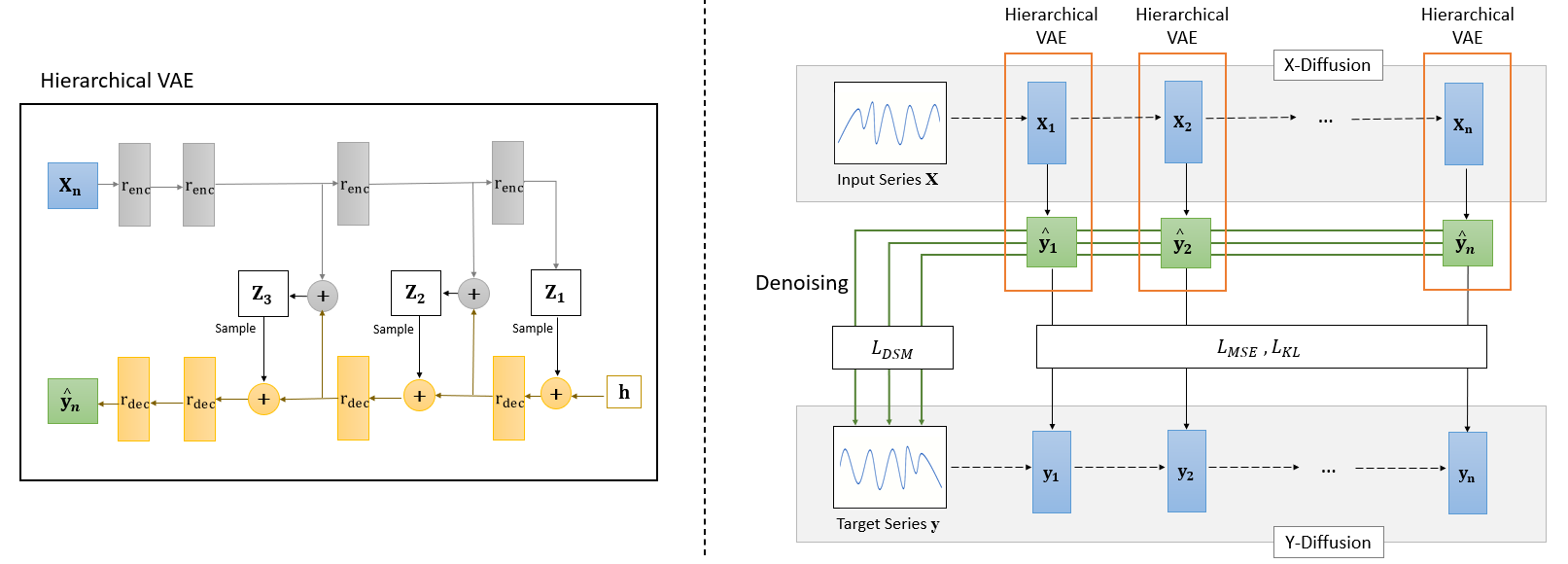}
\vspace{-20px}
  \caption{The overall data generating process of D-Va (right). In D-Va, a diffusion process slowly adds increasing Gaussian noise to both input sequence $\mathbf{X}$ and target sequence $\mathbf{y}$ to emulate the stochastic nature of stock prices. A hierarchical VAE model (left) then learns to generate predictions $\mathbf{\hat{y}}_{n}$ from the noisy inputs $\mathbf{X}_{n}$, which are matched to the noisy targets $\mathbf{y}_{n}$. Finally, we also perform Denoising Score Matching to learn to recover the true target manifold from the noisy predictions (see Figure \ref{target_manifold}).}
\label{model}
\vspace{-8px}
\end{figure*} 
In order to increase the expressiveness of modeling the continuous latent factors that affect stock prices, we leverage deep hierarchical VAEs to learn more complex and low-level latent variables. 

For our backbone model, we use a Noveau Variational Auto-Encoder (NVAE) \cite{vahdat2020nvae} that is repurposed as a seq2seq prediction model. NVAE is a state-of-the-art deep hierarchical VAE model that was originally built for image generation through the use of depthwise separable convolutions and batch normalization. The framework can be seen in Figure \ref{model} (left). In the generative network, a series of decoder residual cells $r_{dec}$, initialized by hidden layer $h$, are trained to generate conditional probability distributions. These are then used to generate the latent variables $\mathbf{Z} = \{\mathbf{Z}_{1}, \mathbf{Z}_{2}, \mathbf{Z}_{3}\}$. The latent variables $\mathbf{Z}$ are further passed on as additional input to the next residual cell, which finally culminates in generating the prediction sequence $\mathbf{\hat{y}}_{n}$. At the same time, in the encoder network, a series of encoder residual cells $r_{enc}$ are used to extract the representations from the input sequence $\mathbf{X}_{n}$, which are also fed to the same generative network to infer the latent variables $\mathbf{Z}$. Formally, we can define the data density of the prediction sequence as:
\vspace{-4px}
\begin{equation}
p(\mathbf{\hat{y}}_{n}|\mathbf{X}_{n})=\int p_{\theta}(\mathbf{Z}|\mathbf{X}_{n})(\mathbf{\hat{y}}_{n}-f(\mathbf{Z}))d\mathbf{Z},
\vspace{-4px}
\end{equation}
where $p_{\theta}$ represents the aggregated data density for all latent variables $\mathbf{Z}$, and $f$ is a parameterized function that represents the aggregated decoder network. The prediction sequence $\mathbf{\hat{y}}_{n}$ can be defined to be generated from the probability distribution $p(\mathbf{\hat{y}}_{n}|\mathbf{X}_{n})$.

\begin{figure}[h]
\includegraphics[width=\linewidth]{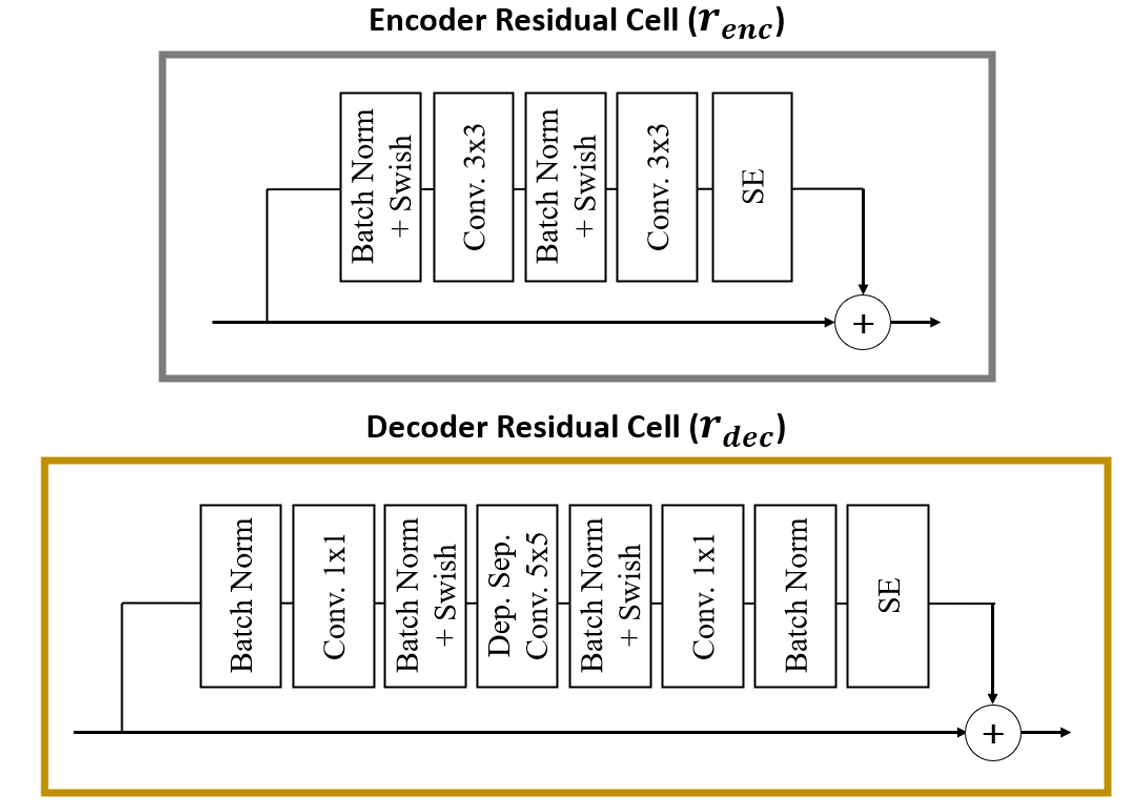}
\vspace{-20px}
 \caption{Breakdown of the encoder/decoder residual cells.}
\label{residual_cells}
\vspace{-15px}
\end{figure}

Next, we follow the work done in \cite{vahdat2020nvae} to design the two types of residual cells, which can be seen in Figure \ref{residual_cells}. The encoder residual cells $r_{enc}$ consist of two series of batch normalization, Swish activation and convolution layers, followed by a Squeeze-and-Excitation (SE) layer. The Swish activation \cite{ramachandran2017searching}, $f(x)=\frac{x}{1+e^{x}}$ and the SE layer \cite{hu2018squeeze}, a gating unit that models the interdependencies between convolutional channels, are both experimentally verified to improve the performance of the hierarchical VAE. For the decoder residual cells $r_{dec}$, a different combination of these units are used, with the addition of a depthwise separable convolution layer. This layer helps to increase the receptive field of the network while keeping computational complexity low by separately mapping all the cross-channels correlations in the input features \cite{chollet2017xception}. This allows us to capture the long-range dependencies of the data within each cell.

The stack of latent variables $\mathbf{Z}$ in the hierarchical VAE, together with the depthwise separable convolution layer in each residual cell, allow us to capture the complex, low-level dependencies of the stock price data beyond its stochasticity. This lets us generate the target sequence more accurately, resulting in better predictions.  
\vspace{-5px}
\subsection{Input Sequence Diffusion}
Next, in order to guide the model in learning from the stochastic stock data, we gradually add random noise to the input stock price sequence, through the use of diffusion probabilistic models.

In the diffusion probabilistic model, we define a Markov chain of diffusion steps that slowly adds random Gaussian noise to the input sequence $\mathbf{X}$ to obtain the noisy samples $\mathbf{X}_{1}, \mathbf{X}_{2},\cdots, \mathbf{X}_{N}$, where $N$ is the number of diffusion steps. The noise to be added at each step is controlled by a variance schedule, $\beta_n\in\left [0, 1\right ]_{n=1}^N$. The noisy samples for each of the diffusion step can then be obtained via:
\vspace{-2px}
\begin{equation}
\begin{gathered}
\begin{aligned}
q(\mathbf{X}_{1:N}|\mathbf{X}) &= \prod^N_{n=1} q(\mathbf{X}_n|\mathbf{X}_{n-1}),  \\
q(\mathbf{X}_n|\mathbf{X}_{n-1}) &= \mathcal{N}(\mathbf{X}_n; \sqrt{1 - \beta_n} \mathbf{X}_{n-1}, \beta_n\mathbf{I});
\end{aligned}
\end{gathered}
\end{equation}
Here, instead of sampling $n$ times for each diffusion step, we use the reparameterization trick \cite{ho2020denoising} to allow us to obtain samples $\mathbf{X}_n$ at any arbitrary step, in order to be able to train the model in a tractable closed-form. Let $\alpha_n = 1 - \beta_n$. We then have:
\begin{equation}
\begin{gathered}
\begin{aligned}
q(\mathbf{X}_n|\mathbf{X}) &= \mathcal{N}(\mathbf{X}_n; \sqrt{\bar{\alpha}_n} \mathbf{X}, (1 - \bar{\alpha}_n)\mathbf{I}), \\
\mathbf{X}_n &= \sqrt{\bar{\alpha}_n} \mathbf{X} + \sqrt{(1 - \bar{\alpha}_n)}\boldsymbol{\epsilon}, \quad\boldsymbol{\epsilon}\sim \mathcal{N}(0,  \textbf{I});
\end{aligned}
\end{gathered}
\end{equation}
This process of generating $\mathbf{X}_{n}$ from $\mathbf{X}$ is akin to performing gradual augmentation on the input series, which trains the model to generate the target sequence through different levels of noise. This then results in more generalizable and robust predictions.
\vspace{-5px}
\subsection{Target Sequence Diffusion}
Additionally, it is shown in in \cite{li2023generative} that by adding diffusion noise to sequences $\mathbf{X}$ and $\mathbf{y}$ concurrently and matching the distributions learnt from a generative model and the diffusion process, it is possible to reduce the overall uncertainty produced by the generative model and the inherent random noise in the data (i.e. the source of aleatoric uncertainty). The relationship can be formulated as:
\begin{equation}
\lim_{n\to\infty}D_{KL}(q(\delta_{\mathbf{\tilde{y}_{n}}})||p(\delta_{\mathbf{\hat{y}_{n}}}|\mathbf{Z}_{n})) < D_{KL}(q(\epsilon_{\mathbf{y}})||p(\epsilon_{\mathbf{\hat{y}}})).
\label{coupled_diffusion}
\end{equation}
Here, the first term is the Kullback–Leibler (KL) divergence between the noise distribution after the additive diffusion process $q(\delta_{\tilde{\mathbf{y}}_{n}})$, and the noise distribution from the generative process of diffused $\mathbf{y}_{n}$ series, $p(\delta_{\hat{\mathbf{y}}_{n}}|\mathbf{Z}_{n})$, \ie the uncertainty after augmentation. The second term is the KL divergence between the inherent data noise distribution $q(\epsilon_{\mathbf{y}})$ and the noise distribution from generating the original $\mathbf{y}$ series, $p(\epsilon_{\hat{\mathbf{y}}})$, \ie the uncertainty before augmentation. Hence, by coupling the generative and diffusion process, the overall prediction uncertainty of the model can be reduced.

Following this observation, we additionally add coupled Gaussian noise to the \textit{target} sequences $\mathbf{y}$ to obtain the noisy samples $\mathbf{y}_{1}, \mathbf{y}_{2},\cdots, \mathbf{y}_{N}$. Here, the noise to be added at each step is $\beta_n' = \gamma\beta_n$ for $\mathbf{y}$, where $\gamma$ is a scaling hyperparameter. Hence, we have:
\vspace{-5px}
\begin{equation}
\begin{gathered}
\begin{aligned}
q(\mathbf{y}_{1:N}|\mathbf{y}) &= \prod^N_{n=1} q(\mathbf{y}_n|\mathbf{y}_{n-1}),  \\
q(\mathbf{y}_n|\mathbf{y}_{n-1}) &= \mathcal{N}(\mathbf{y}_n; \sqrt{1 - \beta_n'} \mathbf{y}_{n-1}, \beta_n'\mathbf{I}).
\end{aligned}
\end{gathered}
\end{equation}
Similarly as before, to allow us to obtain samples $\mathbf{y}_n$ at any arbitrary step, we let $\alpha_n' = 1 - \beta_n'$ and $\bar{\alpha}_n' = \prod_{i=1}^n \alpha_i'$. Then, we have:
\begin{equation}
\begin{gathered}
\begin{aligned}
q(\mathbf{y}_n|\mathbf{y}) &= \mathcal{N}(\mathbf{y}_n; \sqrt{\bar{\alpha}_n'} \mathbf{y}, (1 - \bar{\alpha}_n')\mathbf{I}), \\
\mathbf{y}_n &= \sqrt{\bar{\alpha}_n} \mathbf{y} + \sqrt{(1 - \bar{\alpha}_n)}\boldsymbol{\epsilon}, \quad\boldsymbol{\epsilon}\sim \mathcal{N}(0,  \textbf{I}).
\end{aligned}
\end{gathered}
\end{equation}
For the coupled diffusion process, we minimize the KL divergence:
\begin{equation}
L_{KL} = D_{KL}(p(\mathbf{\hat{y}}_{n})||q(\mathbf{y}_{n})),
\label{KL_loss}
\end{equation}
where $p(\mathbf{\hat{y}}_{n})$ refers to the posterior distribution of the hierarchical VAE that generates the predicted sequence $\mathbf{\hat{y}}_{n}$ at diffusion step $n$, and $q(\mathbf{y}_{n})$ is the corresponding distribution from the diffusion model that generates the noisy target sequence $\mathbf{y}_{n}$. 

Here, the diffusion noise applied to target sequence $\mathbf{y}$ to generate $\mathbf{y}_{n}$ helps to simulate the stochasticity in the target series, similar to what was done for the input sequence $\mathbf{X}$. Additionally, following the theorem in Eqn. \ref{coupled_diffusion}, the coupled diffusion process also allows us to generate the predicted sequence $\mathbf{\hat{y}}$ with less uncertainty. 

\vspace{-5px}
\subsection{Denoising Score-Matching}
In D-Va, the reverse process of a standard diffusion model \cite{ho2020denoising} was replaced by the predictor for $\mathbf{y}$, which removes the need to perform denoising of the diffused samples. At test-time, we simply feed input sequence $\mathbf{X}$ into the hierarchical VAE model, which will predict the target sequence $\mathbf{y}$. However, we note that we have previously defined target sequence $\mathbf{y}$ to be a stochastic series, i.e. $\mathbf{y}=\mathbf{y}_{r}+\epsilon_{\mathbf{y}}$, which is not ideal to recover fully. Instead, we aim to capture the "true" sequence $\mathbf{y}_{r}$, which lies on the actual data manifold. 

\begin{figure}[h]
\vspace{-5px}
\includegraphics[width=\linewidth]{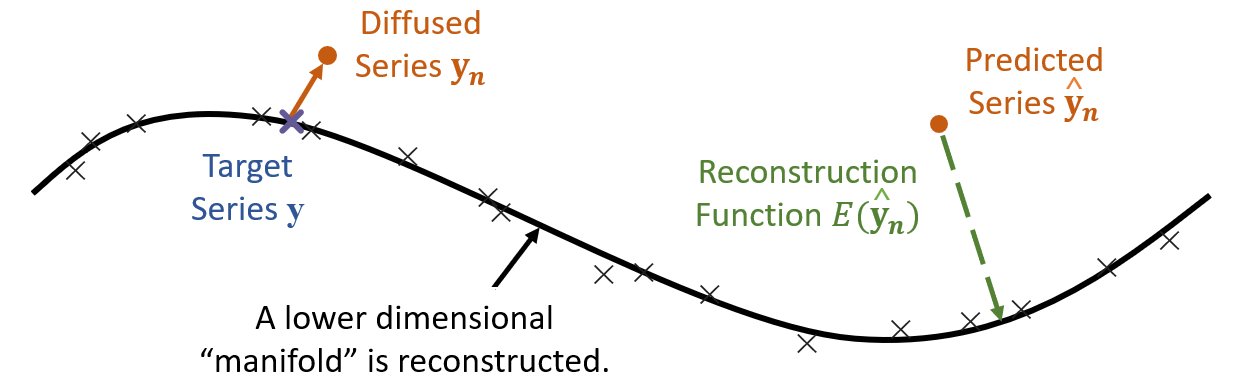}
\vspace{-15px}
 \caption{High-level visualization of the diffusion and denoising process. Note that function $E(\mathbf{\hat{y}}_{n})$ do not reconstruct the target sequence $\mathbf{y}$ exactly, but a lower dimensional "manifold" (black line) that is closer to real sequence $\mathbf{y}_{r}$.}
\label{target_manifold}
\vspace{-14px}
\end{figure}

It has been shown in multiple previous works that it is possible to obtain samples closer to the actual data manifold, by adding an extra denoising step on the final sample \cite{saremi2019neural, kadkhodaie2020solving, song2020improved}. This is akin to helping to remove residual noise, which could be caused by inappropriate sampling steps \cite{jolicoeur2020adversarial}, \etc In our case, this step will serve to remove the intrinsic noise $\epsilon_{\mathbf{y}}$ from the generated target sequence $\mathbf{\hat{y}}$, further reducing the aleatoric uncertainty of the series prediction. To do so, we first follow the denoising score-matching (DSM) process of a standard diffusion probabilistic model \cite{song2019generative, li2019learning}. The process matches the gradient from the noisy prediction $\mathbf{\hat{y}}_{n}$ to the “clean” $\mathbf{y}$ with the gradient of an energy function $\bigtriangledown_{\mathbf{\hat{y}}_{n}}E(\mathbf{\hat{y}}_{n})$ to be learnt, scaled by the amount of noise added from the diffusion process. Note that by doing energy-based learning, the model is not learning to replicate the target y series exactly, but a lower-dimensional manifold, which is closer to the "true" sequence $\mathbf{y}_{r}$ (see Figure \ref{target_manifold}). The DSM loss function to be minimized is as follows: 
\begin{equation}
L_{DSM, n} = \mathbb{E}_{q(\mathbf{\hat{y}}_{n}|\mathbf{y})}\sigma_{n}\left \|\mathbf{y}-\mathbf{\hat{y}}_{n}+\bigtriangledown_{\mathbf{\hat{y}}_{n}}E(\mathbf{\hat{y}}_{n})\right \|^{2}.
\label{DSM_loss}
\end{equation}
The gradient of the learnt energy function $\bigtriangledown_{\mathbf{\hat{y}}_{n}}E(\mathbf{\hat{y}}_{n})$ can be seen as a reconstruction step, which is able to recover $\mathbf{y}_{r}$ from a corrupted $\mathbf{y}$ sequence with any level of Gaussian noise. At test-time, we are then able to perform the one-step denoising jump:
\begin{equation}
\mathbf{\hat{y}}_{final} = \mathbf{\hat{y}} - \bigtriangledown_{\mathbf{\hat{y}}}E(\mathbf{\hat{y}}),
\label{DSM_inference}
\end{equation}
where $\mathbf{\hat{y}}_{final}$ is the final predicted sequence of our model.

Additionally, this one-step denoising process can also be seen as removing the estimated aleatoric uncertainty resulting from data stochasticity \cite{li2023generative}. Here, $\bigtriangledown_{\mathbf{\hat{y}}}E(\mathbf{\hat{y}})$ is an estimation of the sum of the noise produced by the generative VAE and the inherent random noise in the data, which we remove from the prediction series. 

\subsection{Optimization and Inference}
Putting together the loss equations from Eq. \ref{KL_loss} and \ref{DSM_loss}, we get:
\begin{equation}
\mathcal{L} = L_{MSE} + \zeta \cdot L_{KL} + \eta \cdot L_{DSM},
\label{overall_loss}
\end{equation}
where $\zeta$ and $\eta$ refers to the tradeoff parameters. Additionally, $L_{MSE}$ calculate the overall mean squared error (MSE) between the predicted sequence $\mathbf{\hat{y}}_{n}$ and diffused sequence $\mathbf{y}_{n}$ for all diffusion steps $n \in N$, giving us the overall loss function $\mathcal{L}$ for our model. 

The training procedure is as follows: During training, we first apply the coupled diffusion process to both the input and target sequence $\mathbf{X}$ and $\mathbf{y}$ to generate the diffused sequences $\mathbf{X}_{n}$ and $\mathbf{y}_{n}$. We then train the hierarchical VAE to generate predictions $\mathbf{\hat{y}}_{n}$ from the diffused input sequence $\mathbf{X}_{n}$, which are matched to the diffused target sequence $\mathbf{y}_{n}$. Simultaneously, we also train a denoising energy function $E(\mathbf{\hat{y}})$ to obtain the "clean" predictions $\mathbf{\hat{y}}$ from $\mathbf{\hat{y}}_{n}$. 

During inference, the trained hierarchical VAE model is used to generate the predictions $\mathbf{\hat{y}}$ from the input sequences $\mathbf{X}$. The predicted sequences are further "cleaned" by taking a one-step denoising jump (Eqn. \ref{DSM_inference}) to remove the estimated aleatoric uncertainty. This allows us to obtain our final predicted sequences $\mathbf{\hat{y}}_{final}$.

\section{Experiment}
We extensively evaluate D-Va on real-world stock data across three different time periods from 2014-2022, using two different datasets. Our work aims to answer the following three research questions:
\begin{itemize}[leftmargin=*]
  \item \textbf{RQ1}: How does D-Va perform against the state-of-the-art methods on the multi-step regression stock prediction task?

  \item \textbf{RQ2}: How does each of the proposed components (\ie Hierarchical VAE, X-Diffusion, Y-Diffusion, Denoising) affects the prediction performance of D-Va?

  \item \textbf{RQ3}: How does the multi-step outputs of D-Va help institutional investors in a practical setting, \eg portfolio optimization?
\end{itemize}

\vspace{-7px}
\subsection{Experiment Settings}
\subsubsection{\textbf{Dataset}}
The first dataset used is the \textbf{ACL18} StockNet dataset \cite{xu2018stock}. It contains the historical prices of 88 high trade volume stocks from the U.S. market, which represents the top 8-10 stocks in capital size across 9 major industries. The duration of data ranges between 01/01/2014 to 01/01/2017. This dataset is a popular benchmark that has been used in many stock prediction works \cite{feng2021time, sawhney2020deep, feng2018enhancing}.

Furthermore, we extend this dataset by collecting updated U.S. stock data from 01/01/2017 to 01/01/2023, taking the latest top 10 stocks from the 11 major industries (Note that the list of industries have been expanded since the previous work), giving us a total of 110 stocks. The data is collected from Yahoo Finance\footnotemark{}\footnotetext{https://finance.yahoo.com/}, and is processed in the same manner as the \textbf{ACL18} dataset. 

For this work, in order to maintain a consistent dataset length across the experiments, we further split this dataset into two. This gives us exactly three datasets of 3 years each for evaluation. In the results, each dataset will be labelled by their last year, which also correspond to the year of the testing period, \ie 2016, 2019, 2022. 
We summarize the statistics of three datasets in Table \ref{dataset-split}.

For all datasets, we also split all data into training, validation and test sets in chronological order by the ratio of 7:1:2.
\vspace{-8px}
\begin{table}[h]
\caption{Statistics of the datasets.}
\vspace{-10px}
\label{dataset-split}
\begin{tabular}{l|l|l|l}
\hline
Dataset & Duration              & \# Stocks & \# Trading Days \\ \hline\hline
2016    & Jan 01 2014 - Dec 31 2016 & 88        & 756             \\ \hline
2019    & Jan 01 2017 - Dec 31 2019 & 110       & 754             \\ \hline
2022    & Jan 01 2020 - Dec 31 2022 & 110       & 756             \\ \hline
\end{tabular}
\vspace{-15px}
\end{table}
\subsubsection{\textbf{Baselines}}
As the multi-step stock price prediction task has not been widely explored, to demonstrate the effectiveness of D-Va, we also include baselines from the general seq2seq task for comparison. This includes both statistical and deep learning methods that have been shown to work well in time series forecasting.

\definecolor{Gray}{gray}{0.85}
\newcolumntype{a}{>{\columncolor{Gray}}c}
\begin{table*}[h]
\caption{Performance comparisons. The first column shows the test period year and the sequence length $T$. Each result represents the average MSE and standard deviation (in subscript) across 5 runs and all stocks. The best results are boldfaced.}
\label{results}
\vspace{-8px}
\begin{tabular}{l|l|ccccc|acc} 
\hline
\multicolumn{2}{c}{Model}                       & ARIMA                 & NBA                 & VAE                             & VAE + Adv                       & Autoformer                                  & D-Va       & $\uparrow$ MSE & $\uparrow$ SD  \\ 
\hline
\multicolumn{1}{c|}{\multirow{4}{*}{2016}} & 10 & $1.7054_{\pm -}$ & $1.2012_{\pm 0.1190}$ & $1.0743_{\pm 0.0255}$             & $1.0752_{\pm 0.0249}$             & $\underline{1.0204}_{\pm \underline{0.0179}}$ & $\mathbf{0.9040_{\pm 0.0048}}$ & 11.41\%                            & 73.06\%        \\
\multicolumn{1}{c|}{}                      & 20 & $1.5918_{\pm -}$ & $1.1656_{\pm 0.1024}$ & $1.0652_{\pm 0.0223}$             & $1.0639_{\pm 0.0202}$             & $\underline{0.9920}_{\pm \underline{0.0197}}$ & $\mathbf{0.9144_{\pm 0.0047}}$ & 7.83\%                             & 76.16\%        \\
\multicolumn{1}{c|}{}                      & 40 & $0.9135_{\pm -}$ & $0.9713_{\pm 0.0742}$ & $0.9918_{\pm \underline{0.0159}}$ & $0.9920_{\pm 0.0177}$             & $\underline{0.8860}_{\pm 0.0241}$             & $\mathbf{0.8591_{\pm 0.0052}}$ & 3.04\%                             & 67.56\%        \\
\multicolumn{1}{c|}{}                      & 60 & $\underline{0.8241}_{\pm -}$ & $0.8909_{\pm 0.0373}$ & $0.9390_{\pm \underline{0.0173}}$ & $0.9397_{\pm 0.0178}$             & $0.8625_{\pm 0.0378}$             & $\mathbf{0.8035_{\pm 0.0050}}$ & 2.50\%                             & 71.09\%        \\ 
\hline
\multirow{4}{*}{2019}                      & 10 & $ 1.6737_{\pm -}$ & $1.2266_{\pm 0.1070}$ & $1.1527_{\pm 0.0271}$             & $1.1524_{\pm 0.0250}$             & $\underline{1.1011}_{\pm \underline{0.0180}}$ & $\mathbf{0.9847_{\pm 0.0040}}$ & 10.57\%                            & 77.87\%        \\
                                           & 20 & $1.2115_{\pm -}$ & $1.2166_{\pm 0.1161}$ & $1.1156_{\pm \underline{0.0186}}$ & $1.1150_{\pm 0.0192}$             & $\underline{1.0762}_{\pm 0.0224}$             & $\mathbf{0.9703_{\pm 0.0039}}$ & 9.84\%                             & 79.10\%        \\
                                           & 40 & $\underline{1.1189}_{\pm -}$ & $1.1785_{\pm 0.0758}$ & $1.1736_{\pm 0.0168}$             & $1.1744_{\pm \underline{0.0163}}$ & $1.1371_{\pm 0.0339}$             & $\mathbf{1.0453_{\pm 0.0040}}$ & 6.58\%                             & 75.40\%        \\
                                           & 60 & $\underline{1.0887}_{\pm -}$ & $1.1395_{\pm 0.0545}$ & $1.1841_{\pm 0.0183}$             & $1.1830_{\pm \underline{0.0180}}$ & $1.2066_{\pm 0.0501}$             & $\mathbf{1.0505_{\pm 0.0043}}$ & 3.51\%                            & 75.93\%        \\ 
\hline
\multirow{4}{*}{2022}                      & 10 & $1.2857_{\pm -}$ & $0.9154_{\pm 0.0261}$ & $1.0297_{\pm 0.0223}$             & $1.0196_{\pm 0.0215}$             & $\underline{0.9958}_{\pm \underline{0.0179}}$ & $\mathbf{0.8653_{\pm 0.0053}}$ & 13.10\%                            & 70.23\%        \\
                                           & 20 & $1.0095_{\pm -}$ & $0.9346_{\pm 0.0455}$ & $0.9810_{\pm \underline{0.0171}}$ & $0.9811_{\pm 0.0179}$             & $\underline{0.9529}_{\pm 0.0207}$             & $\mathbf{0.8451_{\pm 0.0037}}$ & 10.70\%                            & 78.41\%        \\
                                           & 40 & $0.9384_{\pm -}$ & $0.8739_{\pm 0.0301}$ & $0.9643_{\pm 0.0156}$             & $0.9635_{\pm \underline{0.0155}}$ & $\underline{0.9142}_{\pm 0.0227}$             & $\mathbf{0.8426_{\pm 0.0035}}$ & 7.83\%                             & 77.12\%        \\
                                           & 60 & $0.8538_{\pm -}$ & $0.8625_{\pm 0.0195}$ & $0.9436_{\pm 0.0171}$             & $0.9442_{\pm \underline{0.0162}}$ & $\underline{0.8428}_{\pm 0.0254}$             & $\mathbf{0.8174_{\pm 0.0035}}$ & 3.01\%                             & 78.22\%        \\
\hline
\end{tabular}
\vspace{-10px}
\end{table*}

\begin{itemize}[leftmargin=*]

  \item \textbf{ARIMA} \cite{box2015time} Autoregressive Integrated Moving Average (ARIMA) is a traditional statistical method that combines autoregressive, differencing, and moving average components to do time series forecasting. This is one of the baselines in the M5 accuracy competition, a popular time series forecasting competition \cite{makridakis2022m5}.
  
  \item \textbf{NBA} \cite{liu2018numerical}: Numerical-Based Attention (NBA) is a baseline model that tackles the multi-step stock price prediction task. It utilizes a long short-term memory (LSTM) network \cite{hochreiter1997long, akita2016deep} with an additional attention component that captures the text and temporal dependency of stock prices. For this model, we remove the text input component for an equivalent comparison.

  \item \textbf{VAE} \cite{kingma2013auto, xu2018stock} We adapt a vanilla VAE model as a benchmark to compare against the hierarchical VAE. In this model, there is a single dense layer for the encoder, a single latent variable to be generated and a single dense layer for the sampling decoder. The VAE model has been shown to provide improvements in the single-step stock movement classification task \cite{xu2018stock}.

  \item \textbf{VAE + Adversarial} \cite{goodfellow2014explaining, feng2018enhancing}: This is similar to the above model, but adversarial perturbations are added to the input sequence. The perturbation added is equal to the gradient of the loss function of the model, similar to what was done in \cite{goodfellow2014explaining, feng2018enhancing}. 

  \item \textbf{Autoformer} \cite{wu2021autoformer}: The Autoformer is a state-of-the-art seq2seq prediction model that aggregates series-wise dependencies to do long-term forecasting. The model adapts a standard Transformer model with progressive decomposition capacities and an auto-correlation mechanism to find series periodicity. 
\end{itemize}
\vspace{-20px}

\subsubsection{\textbf{Parameter Settings}}
To compare performances under different horizons, we evaluate a range of sequence lengths, with 10, 20, 40 and 60 days for both the input and output lengths $T$ and $T'$. The chosen lengths are typical liquidity horizons required by financial regulators, which follows the motivation of this work.

The Adam optimizer \cite{kingma2014adam} was used to optimize the model, with an initial learning rate of 5e-4. The batch size is 16, and we train the model for 20 epochs for each stock separately, taking the iteration with the best validation results for each. For the trade-off hyperparameters, we perform a grid search of step 0.1 within the range of $\left [0.1, 1 \right ]$ and set $\zeta=0.5$ and $\eta=1$. All experiments are repeated five times for each stock: we report the overall average MSE and standard deviation of all stocks in our results section. 
We then measure the percentage improvement in MSE and standard deviation of D-Va against the strongest baselines for each experiment setting.
\vspace{-20px}
\section{Results}
Next, we will discuss the performance of D-Va in tackling each of the proposed research questions from the previous section.

\subsection{Performance Comparison (RQ1)}
Table \ref{results} reports the results on the multi-step regression stock prediction task. From the table, we observe the following:
\begin{itemize}[leftmargin=*]
\vspace{-5px}
  \item The statistical model ARIMA tends to perform better than the other baselines when the sequence length $T$ is longer. This can be attributed to it having enough information to capture the auto-correlations within the sequence for forecasting, as opposed to shorter sequences where there are less obvious periodicity and greater noise-to-data ratio. This is also later observed in the Autoformer, which also learns from sequence auto-correlations.
  
  \item The VAE models tend to exhibit MSE improvements when the prediction uncertainty is high, as measured by the standard deviation over multiple runs. This can be observed when comparing the performance of the NBA and VAE model. When the predictions of the NBA model display high standard deviation, \ie above 0.075, the VAE model show clear MSE improvements. On the other hand, when the standard deviation for NBA is low, the VAE model performs worse. In all cases, the standard deviation of the VAE model predictions is much lower. It is likely that the NBA model is unable to learn well when the data is noisy, resulting in predictions that are poor and vary greatly. However, during periods where the data is less noisy, the LSTM + Attention components of the NBA model is able to perform better than the simple dense layers of the encoder/decoder in the VAE models. 

  \item The VAE + Adversarial model does not show much improvements from the VAE model. The difference in MSE is not statistically significant, and could be attributed to the standard deviation of the results. However, the additional adversarial component seems to provide some slight improvements in the standard deviation of its prediction results over the pure VAE model. 

  \item The Autoformer model remarkably outperforms the above three deep learning models across all sequence lengths despite not having a variational component, which highlights the robustness of this method. This trait was also mentioned in the source material \cite{wu2021autoformer}, where the model was able to perform well on the exchange rate prediction task even when there are no obvious periodicity. Our result is able to verify this on the stock prediction task. 
  We also observe that the standard deviation of the predictions of this model increases gradually as the sequence length increases, which goes in an opposite trend to the previous 3 models. This could be attributed to noise accumulating in the captured series periodicity which is used to generate the output sequence. 

  \item Our D-Va model is able to outperform all models in terms of its MSE performance and the standard deviation of the predictions. On average, D-Va achieves an improvement of 7.49\% over the strongest baselines (underlined in Table \ref{results}) on MSE performance and a strong 75.01\% on standard deviation. This showcases the capabilities of D-Va in handling data uncertainty to improve predictions in the multi-step regression stock prediction task. 
  
  \item Finally, we note that the MSE improvement decreases with increasing prediction length. This is likely due to the fact that there are more chances of unexpected material changes in the market with a longer prediction horizon, which the model cannot foresee. In a practical setting, given a long horizon, it could be better to do a rolling forecast using a shorter-length model instead.
\end{itemize}
\vspace{-5px}

\subsection{Model Study (RQ2)}
To demonstrate the effectiveness of each additional component in D-Va, we conduct an ablation study over different variants of the model. We remove one additional component for each variant, \ie no denoising component \textbf{(D-Va$-$Dn)}; no target series diffusion and denoising components \textbf{(D-Va$-$YdDn)}; and no input series diffusion, target series diffusion and denoising components, which leaves only the backbone hierarchical VAE model \textbf{(D-Va$-$XdYdDn)}.
\vspace{-5px}
\subsubsection{Ablation Study}
\vspace{-10px}
\begin{table}[h]\small
\caption{Ablation study. The first column shows the test period year and the sequence length $T$. Each result represents the average MSE and standard deviation (in subscript) across 5 runs and all stocks. The best results are boldfaced.}
\label{ablation-results}
\vspace{-7px}
\begin{tabular}{cccccc}
\hline
\multicolumn{2}{c}{Model}                                            & \begin{tabular}[c]{@{}c@{}}D-Va$-$\\ XdYdDn\end{tabular}                  & \begin{tabular}[c]{@{}c@{}}D-Va$-$\\ YdDn\end{tabular}                             & \begin{tabular}[c]{@{}c@{}}D-Va$-$\\ Dn\end{tabular}                               & D-Va                           \\ \hline
\multicolumn{1}{c|}{\multirow{4}{*}{ \rotatebox[origin=c]{90}{2016}}} & \multicolumn{1}{c|}{\rotatebox[origin=c]{90}{10}} & $0.9105_{\pm 0.0058}$          & $0.9040_{\pm 0.0050}$                   & $0.9058_{\pm \textbf{0.0044}}$          & $\textbf{0.9040}_{\pm 0.0048}$ \\
\multicolumn{1}{c|}{}                      & \multicolumn{1}{c|}{\rotatebox[origin=c]{90}{20}} & $0.9266_{\pm 0.0079}$          & $0.9153_{\pm 0.0052}$                   & $0.9161_{\pm \textbf{0.0035}}$          & $\textbf{0.9144}_{\pm 0.0047}$ \\
\multicolumn{1}{c|}{}                      & \multicolumn{1}{c|}{\rotatebox[origin=c]{90}{40}} & $0.8737_{\pm 0.0103}$          & $0.8595_{\pm 0.0047}$                   & $0.8592_{\pm \textbf{0.0029}}$          & $\textbf{0.8591}_{\pm 0.0052}$ \\
\multicolumn{1}{c|}{}                      & \multicolumn{1}{c|}{\rotatebox[origin=c]{90}{60}} & $0.8157_{\pm 0.0108}$          & $\textbf{0.8030}_{\pm 0.0045}$          & $0.8034_{\pm \textbf{0.0030}}$          & $0.8035_{\pm 0.0050}$          \\ \hline
\multicolumn{1}{c|}{\multirow{4}{*}{\rotatebox[origin=c]{90}{2019}}} & \multicolumn{1}{c|}{\rotatebox[origin=c]{90}{10}} & $0.9885_{\pm 0.0049}$          & $0.9857_{\pm 0.0049}$                   & $0.9856_{\pm \textbf{0.0039}}$          & $\textbf{0.9847}_{\pm 0.0040}$ \\
\multicolumn{1}{c|}{}                      & \multicolumn{1}{c|}{\rotatebox[origin=c]{90}{20}} & $0.9763_{\pm 0.0059}$          & $0.9708_{\pm 0.0039}$                   & $0.9707_{\pm \textbf{0.0031}}$          & $\textbf{0.9703}_{\pm 0.0039}$ \\
\multicolumn{1}{c|}{}                      & \multicolumn{1}{c|}{\rotatebox[origin=c]{90}{40}} & $1.0576_{\pm 0.0094}$          & $1.0451_{\pm 0.0037}$                   & $\textbf{1.0449}_{\pm \textbf{0.0027}}$ & $1.0453_{\pm 0.0040}$          \\
\multicolumn{1}{c|}{}                      & \multicolumn{1}{c|}{\rotatebox[origin=c]{90}{60}} & $1.0622_{\pm 0.0105}$          & $1.0496_{\pm 0.0036}$                   & $\textbf{1.0496}_{\pm \textbf{0.0025}}$ & $1.0505_{\pm 0.0043}$          \\ \hline
\multicolumn{1}{c|}{\multirow{4}{*}{\rotatebox[origin=c]{90}{2022}}} & \multicolumn{1}{c|}{\rotatebox[origin=c]{90}{10}} & $\textbf{0.8638}_{\pm 0.0039}$ & $0.8643_{\pm 0.0040}$                   & $0.8646_{\pm \textbf{0.0034}}$          & $0.8653_{\pm 0.0053}$          \\
\multicolumn{1}{c|}{}                      & \multicolumn{1}{c|}{\rotatebox[origin=c]{90}{20}} & $0.8466_{\pm 0.0043}$          & $\textbf{0.8450}_{\pm 0.0036}$          & $0.8452_{\pm \textbf{0.0026}}$          & $0.8451_{\pm 0.0037}$          \\
\multicolumn{1}{c|}{}                      & \multicolumn{1}{c|}{\rotatebox[origin=c]{90}{40}} & $0.8469_{\pm 0.0064}$          & $\textbf{0.8421}_{\pm 0.0031}$          & $0.8427_{\pm \textbf{0.0027}}$          & $0.8426_{\pm 0.0035}$          \\
\multicolumn{1}{c|}{}                      & \multicolumn{1}{c|}{\rotatebox[origin=c]{90}{60}} & $0.8254_{\pm 0.0081}$          & $\textbf{0.8162}_{\pm 0.0030}$ & $0.8172_{\pm \textbf{0.0029}}$                   & $0.8174_{\pm 0.0035}$          \\ \hline
\end{tabular}
\vspace{-10px}
\end{table}

Table \ref{ablation-results} reports the results on the ablation study. From the table, we make the following observations:
\begin{itemize}[leftmargin=*]
  \item The backbone model \textbf{D-Va$-$XdYdDn}, \ie only the hierarchical VAE, already showcase a strong MSE improvement over the previous baseline models, \ie ARIMA and Autoformer, which highlights the strength of this method. It is possible that handling the data stochasticity through the latent variables of the hierarchical VAE is more effective than learning from the series auto-correlation, for the stock price prediction task. 



  \item With each additional diffusion component, \ie in the \textbf{D-Va$-$YdDn} and \textbf{D-Va$-$Dn} models, we see can clear improvement in the standard deviation of the predictions. The additional noise augmentation helps to improve the stability of the predictions, which was also previously observed in the VAE + Adversarial model. This also shows that our proposed components help to increase the robustness of the model and decrease the prediction uncertainty. 

  \item The best MSE performances seem to vary across the variant models. However, we can see a visible relationship between the prediction uncertainty of one model, as measured by the standard deviation of their results, against the MSE performance of the next model. When the standard deviation is low, \ie below 0.050, there seems to be little to no MSE improvement provided by the next additional component. It is likely that the components work by making the model more robust against noise in the data -- however, for experimental settings where the data noise is low enough (hence, low standard deviation), there is not much improvements to be made in the predictions.  We will explore this observation in more details in the next subsection.

  \item Interestingly, we note that the denoising component, \ie from \textbf{D-Va$-$Dn} to \textbf{D-Va}, also slightly increase the standard deviation of the predictions. This could be due to it being trained to take the one-step denoising step \textit{towards} the target series $\mathbf{y}$ (see Eq. \ref{DSM_loss}), which we had previously defined to be stochastic. 
\end{itemize}

\vspace{-5px}
\subsubsection{Uncertainty Reduction}
A key observation that was made over the results analysis was that the MSE improvements of the variational components seem to be related to the uncertainty, measured by the standard deviation of the results, from the prior models. In this section, we study this observation in more detail. 

First, we note that the reported performance was calculated from the average of the results from individual stocks, each of them with their own average MSE and standard deviation over 5 runs. 
Taking all of their individual results from all 12 experimental settings (\ie 3 test periods and 4 sequence lengths), we then plot the standard deviation for one model against the percentage change in MSE from the improved variant, given an additional component. 

\vspace{-5px}
\begin{figure}[h]
\includegraphics[width=\linewidth]{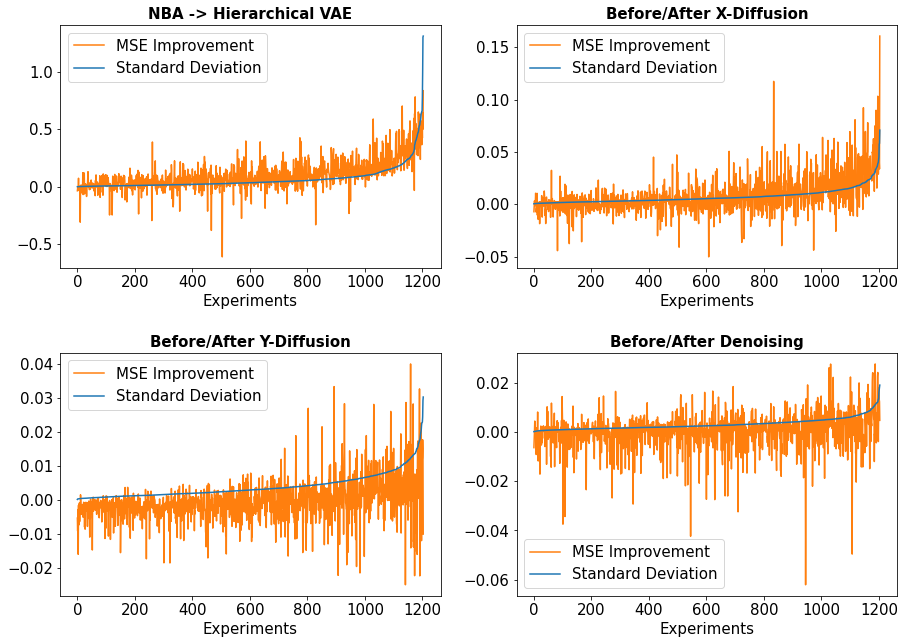}
\vspace{-21px}
 \caption{Relationship between standard deviation of the prediction results from an initial model and the percentage improvement in MSE from the additional components. Both variables share the same scale on the y-axis. In each subplot, the experimental results are sorted by the standard deviation variable from lowest on the left to highest on the right.}
\label{stddev_mse}
\vspace{-9px}
\end{figure} 
As shown in Figure \ref{stddev_mse}, there is a visible relationship between the prediction uncertainty of the models, as measured by the standard deviation of the prediction results before the introduction of each variational component, with the resulting MSE improvement after each component is introduced. When the model is more uncertain about its predictions, likely due to the stochastic nature of stock data, the additional VAE and diffusion components become more effective given their capabilities to handle data noise and uncertainty. The relationship becomes less pronounced with each additional component, likely due to there being significantly less uncertainty with each component that comes before it (note the reduction in scale of the standard deviation across each subplot in Figure \ref{stddev_mse}).

This relationship is least visible in the denoising component. This could be attributed to two possible reasons: Firstly, through the VAE and diffusion components, the data noise has been reduced to the minimal possible, making it difficult for the last component to produce any more improvements. Secondly, it is also plausible that the denoising component works via a different mechanism that does not depend as much on the prior model's uncertainty. As mentioned, the model is trained to take the one-step denoising step \textit{towards} the stochastic target series $\mathbf{y}$, which increases its information of the targets but reduces its generalizability. This is similar to the bias-variance problem \cite{kohavi1996bias, von2011statistical}, whereas our model does not want to overfit to the target data as it also contains stochastic noise.

\subsection{Portfolio Optimization (RQ3)}
Furthermore, we examine the usefulness of the model in a practical setting. As opposed to a single-day prediction, having multi-step prediction allow us to understand the outlook for the stocks' returns over the next few days, allowing us to form a portfolio of assets that can maximize our returns and minimize its volatility. 

\subsubsection{Mean-Variance Optimization}
Given the returns sequence prediction $\mathbf{\hat{y}}_{s}$ for every stock $s$, we first calculate their overall mean and covariance across each individual prediction period $t$:
\begin{equation}
\begin{gathered}
\boldsymbol{\mu}_{t} = AVG(\mathbf{\hat{y}}_{1, t}, \mathbf{\hat{y}}_{2, t}, \cdots,  \mathbf{\hat{y}}_{s, t}), \\
\boldsymbol{\Sigma}_{t} = COV(\mathbf{\hat{y}}_{1, t}, \mathbf{\hat{y}}_{2, t}, \cdots, \mathbf{\hat{y}}_{s, t})  ,
\end{gathered}
\end{equation}
where $\boldsymbol{\mu}_{t} \in \mathbb{R}^{S}$ and $\boldsymbol{\Sigma}_{t} \in \mathbb{R}^{S\times S}$ represents the mean vector and covariance matrix of all stocks' returns across prediction period $t$, and $S$ is the total number of stocks. We then form a stock portfolio for the period using Markowitz' mean-variance optimization \cite{selection1952harry}:
\vspace{-2px}
\begin{equation}
\begin{gathered}
\underset{\mathbf{\mathit{w}}}{\textup{max}}\; \mathbf{\mathit{w'}}\mathbf{\mu}-\frac{\gamma}{2}\mathbf{\mathit{w'}}\mathbf{\Sigma}\mathbf{\mathit{w}}, \vspace{-3px}\\
\textup{s.t.}\;\mathbf{\mathit{w'}}\mathbf{1} = 1,
\end{gathered}
\label{portfolio_optimization}
\vspace{-2px}
\end{equation}
where $\mathbf{\mathit{w}}$ is the portfolio weights vector to be learnt, which sums to 1 representing the overall capital. $\gamma$ is the risk-aversion parameter, which can be treated as a hyper-parameter to be tuned by maximizing the portfolio results on the validation set \cite{peng2022portfolio}. Additionally, we also set a no short-sales constraint, \ie $\mathit{w}_{s} \geq 0 \; (s = 1, 2, \cdots, S)$, which has been shown to reduce the overall portfolio risk \cite{jagannathan2003risk, demiguel2009optimal} and is also often restricted by financial institutions \cite{chang2014short, beber2013short}. 

\begin{table*}[]\footnotesize
\caption{Comparison of 10-Day Sharpe ratios. For each table, going from left to right represents the handling of epistemic uncertainty and going from top to bottom represents the handling of aleatoric uncertainty. The best results are boldfaced.}
\label{sharpe_results}
\vspace{-5px}
\begin{tabular}{cccllcccllcccl}
\cline{1-3} \cline{6-8} \cline{11-13}
\multicolumn{1}{|c|}{\normalsize{2016}}                  & \multicolumn{1}{c|}{Sharpe Ratio}         & \multicolumn{1}{c|}{\begin{tabular}[c]{@{}c@{}}Sharpe Ratio \\ (Regularized)\end{tabular}} &                                    & \multicolumn{1}{l|}{} & \multicolumn{1}{c|}{\normalsize{2019}}                  & \multicolumn{1}{c|}{Sharpe Ratio}         & \multicolumn{1}{c|}{\begin{tabular}[c]{@{}c@{}}Sharpe Ratio \\ (Regularized)\end{tabular}} &                                    & \multicolumn{1}{l|}{} & \multicolumn{1}{c|}{\normalsize{2022}}                  & \multicolumn{1}{c|}{Sharpe Ratio}         & \multicolumn{1}{c|}{\begin{tabular}[c]{@{}c@{}}Sharpe Ratio \\ (Regularized)\end{tabular}} &                                    \\ \cline{1-3} \cline{6-8} \cline{11-13}
\multicolumn{1}{|c|}{\multirow{2}{*}{NBA}}   & \multicolumn{1}{c|}{\multirow{2}{*}{\normalsize{0.0270}}} & \multicolumn{1}{c|}{\multirow{2}{*}{\normalsize{0.0691}}}                                                           & \multirow{6}{*}{ \rotatebox[origin=c]{270}{\tiny{Handling aleatoric $\rightarrow$ }}} & \multicolumn{1}{l|}{} & \multicolumn{1}{c|}{\multirow{2}{*}{NBA}}   & \multicolumn{1}{c|}{\multirow{2}{*}{\normalsize{0.0820}}} & \multicolumn{1}{c|}{\multirow{2}{*}{\normalsize{0.1767}}}                                                           & \multirow{6}{*}{ \rotatebox[origin=c]{270}{\tiny{Handling aleatoric $\rightarrow$ }}} & \multicolumn{1}{l|}{} & \multicolumn{1}{c|}{\multirow{2}{*}{NBA}}   & \multicolumn{1}{c|}{\multirow{2}{*}{\normalsize{0.0332}}} & \multicolumn{1}{c|}{\multirow{2}{*}{\normalsize{0.0404}}}                                                           & \multirow{6}{*}{ \rotatebox[origin=c]{270}{\tiny{Handling aleatoric $\rightarrow$ }}} \\
\multicolumn{1}{|c|}{}                       & \multicolumn{1}{c|}{}                       & \multicolumn{1}{c|}{}                                                                                 &                                    & \multicolumn{1}{l|}{} & \multicolumn{1}{c|}{}                       & \multicolumn{1}{c|}{}                       & \multicolumn{1}{c|}{}                                                                                 &                                    & \multicolumn{1}{l|}{} & \multicolumn{1}{c|}{}                       & \multicolumn{1}{c|}{}                       & \multicolumn{1}{c|}{}                                                                                 &                                    \\ \cline{1-3} \cline{6-8} \cline{11-13}
\multicolumn{1}{|c|}{\multirow{2}{*}{Equal}} & \multicolumn{2}{c|}{\multirow{2}{*}{\normalsize{0.1089}}}                                                                                                         &                                    & \multicolumn{1}{l|}{} & \multicolumn{1}{c|}{\multirow{2}{*}{Equal}} & \multicolumn{2}{c|}{\multirow{2}{*}{\normalsize{0.2337}}}                                                                                                         &                                    & \multicolumn{1}{l|}{} & \multicolumn{1}{c|}{\multirow{2}{*}{Equal}} & \multicolumn{2}{c|}{\multirow{2}{*}{\normalsize{0.0437}}}                                                                                                         &                                    \\
\multicolumn{1}{|c|}{}                       & \multicolumn{2}{c|}{}                                                                                                                               &                                    & \multicolumn{1}{l|}{} & \multicolumn{1}{c|}{}                       & \multicolumn{2}{c|}{}                                                                                                                               &                                    & \multicolumn{1}{l|}{} & \multicolumn{1}{c|}{}                       & \multicolumn{2}{c|}{}                                                                                                                               &                                    \\ \cline{1-3} \cline{6-8} \cline{11-13}
\multicolumn{1}{|c|}{\multirow{2}{*}{D-Va}}  & \multicolumn{1}{c|}{\multirow{2}{*}{\normalsize{0.0772}}} & \multicolumn{1}{c|}{\multirow{2}{*}{\textbf{\normalsize{0.1174}}}}                                                  &                                    & \multicolumn{1}{l|}{} & \multicolumn{1}{c|}{\multirow{2}{*}{D-Va}}  & \multicolumn{1}{c|}{\multirow{2}{*}{\normalsize{0.1197}}} & \multicolumn{1}{c|}{\multirow{2}{*}{\textbf{\normalsize{0.2767}}}}                                                  &                                    & \multicolumn{1}{l|}{} & \multicolumn{1}{c|}{\multirow{2}{*}{D-Va}}  & \multicolumn{1}{c|}{\multirow{2}{*}{\normalsize{0.0600}}} & \multicolumn{1}{c|}{\multirow{2}{*}{\textbf{\normalsize{0.0645}}}}                                                  &                                    \\
\multicolumn{1}{|c|}{}                       & \multicolumn{1}{c|}{}                       & \multicolumn{1}{c|}{}                                                                                 &                                    & \multicolumn{1}{l|}{} & \multicolumn{1}{c|}{}                       & \multicolumn{1}{c|}{}                       & \multicolumn{1}{c|}{}                                                                                 &                                    & \multicolumn{1}{l|}{} & \multicolumn{1}{c|}{}                       & \multicolumn{1}{c|}{}                       & \multicolumn{1}{c|}{}                                                                                 &                                    \\ \cline{1-3} \cline{6-8} \cline{11-13}
\multicolumn{1}{l}{}                         & \multicolumn{2}{c}{\tiny{Handling epistemic $\longrightarrow$}}                                                                                                               &                                    &                       & \multicolumn{1}{l}{}                        & \multicolumn{2}{c}{\tiny{Handling epistemic $\longrightarrow$}}                                                                                                               & \multicolumn{1}{c}{}               &                       & \multicolumn{1}{l}{}                        & \multicolumn{2}{c}{\tiny{Handling epistemic $\longrightarrow$}}                                                                                                               & \multicolumn{1}{c}{}              
\end{tabular}
\vspace{-10px}
\end{table*}
\subsubsection{Graphical Lasso Regularization}
Additionally, we further perform regularization on the covariance matrix by applying a graphical lasso \cite{friedman2008sparse}. This is done by maximizing the penalized log-likelihood:
\begin{equation}
\underset{\Theta}{\textup{max}}\:\textup{log}\:\textup{det} \Theta-\textup{tr}(\Sigma\Theta) - \lambda\left \| \Theta \right \|_{1}
\vspace{-2px}
\end{equation}
where $\Theta$ is the regularized, \textit{inverted} covariance matrix to be learnt and $\lambda$ is a hyper-parameter, which we set to 0.1 in our experiments. The method is akin to performing L1 regularization \cite{tibshirani1996regression} in machine learning, which increases the generalizability of the predictions by reducing their dimensionalities. It is also similar to performing covariance shrinkage \cite{ledoit2004honey, ledoit2004well} in finance, where the extreme values are pulled towards more central values. However, the graphical lasso has been shown to work better for the covariances of smaller samples \cite{belilovsky2017learning, gao2018modeling}, which was also observed in our experiments.

The L1 regularization can also be seen as a way to reduce the impact of model uncertainty, which is a form of epistemic uncertainty \cite{hullermeier2021aleatoric}. On the other hand, our diffusion-based prediction model deals mainly with data uncertainty, or aleatoric uncertainty \cite{li2023generative}. Applying both techniques, we explore the effects of handling each type of uncertainty on the performance of the stock portfolio.

\vspace{-6px}
\subsubsection{Comparison Method}
To compare portfolio performance, we use the Sharpe Ratio \cite{sharpe1998sharpe}, which is a measure of the ratio of the portfolio returns compared to its volatility. It is defined as the overall expected returns of the portfolio $\hat{\mu}$ (we set the risk-free interest rate to 0) divided by the returns' standard deviation $\hat{\sigma}$, \ie $SR = \frac{\hat{\mu}}{\hat{\sigma}}$:

For each prediction period $t$, we first form a portfolio across sequence length $T$ using Eq. \ref{portfolio_optimization}, which allows us to calculate its Sharpe Ratio. We then evaluate the average Sharpe Ratio across all $t$ within the test period, and also average across the results of the 5 prediction runs.
Additionally, we also include the performance of the equal-weight portfolio, \ie $\mathit{w}_{s} = \frac{1}{S}$. The naive equal-weight portfolio have been shown to outperform most existing portfolio methods on look-ahead performance \cite{demiguel2009optimal}, which makes it a strong baseline for comparison. We make three comparisons: Firstly, we compare the average 10-day Sharpe ratios of D-Va with the baseline model NBA, with and without using regularization, to analyze the impact of handling each uncertainty type. Next, we compare the average $T$-day Sharpe ratios across the different sequence lengths $T$.
Finally, we compare the average $T$-day Sharpe ratios, after regularization, across the benchmark multi-step prediction models.

\vspace{-6px}
\subsubsection{Portfolio Results}
Table \ref{sharpe_results} compares the average 10-day Sharpe ratios across the different dataset years. We can see that using D-Va as the prediction model and applying regularization both help to improve the Sharpe ratio results, and combining both methods allow us to obtain the best Sharpe ratio performance. The equal-weight portfolio remains a strong baseline against our non-regularized model: this could be due to D-Va not incorporating additional information sources like news, and hence not being able to expect new changes or shocks to the price trends. However, the model was still able to provide enough information to the regularized portfolio method to outperform this benchmark using only historical price data, which is an optimistic sign for its prediction ability.

\begin{figure}[t]
\includegraphics[width=\linewidth]{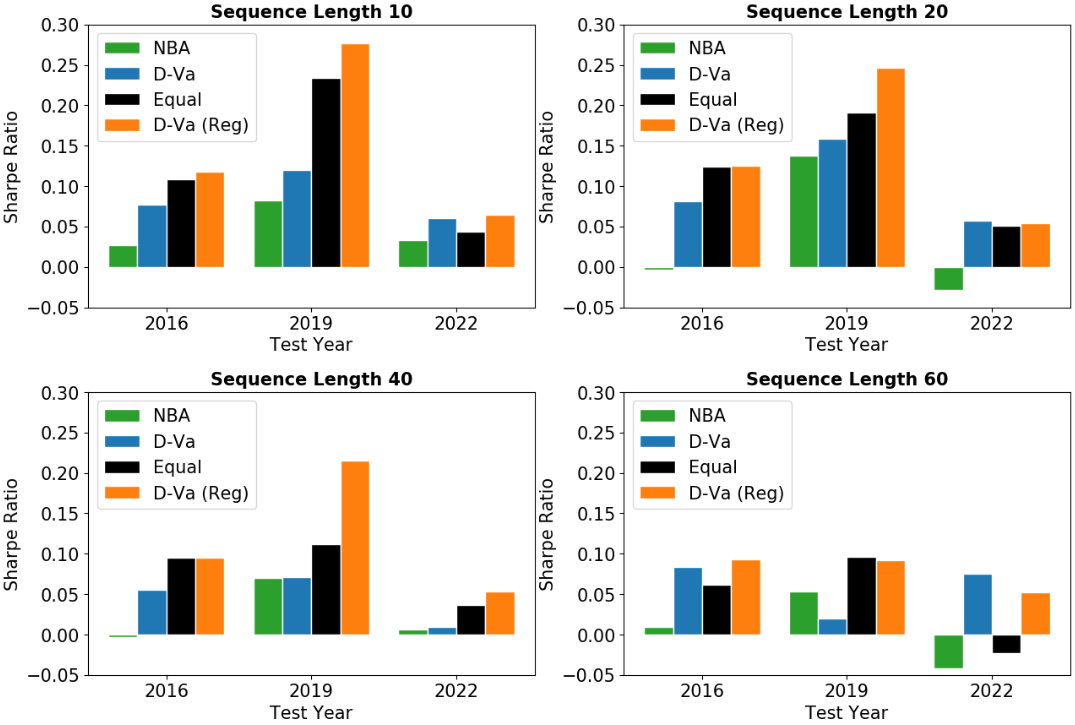}
\vspace{-20px}
 \caption{Comparison of $T$-day Sharpe ratios across the different prediction lengths $T$.}
\label{sharpe_cmp}
\vspace{-16px}
\end{figure}

Figure \ref{sharpe_cmp} compares the average $T$-day Sharpe ratios across the different prediction lengths. As observed previously, using the predictions from D-Va, together with covariance regularization, we are able to form portfolios that consistently achieve the best Sharpe ratios, even across difference sequence lengths $T$. We note that there are settings where D-Va with regularization does not outperform the equal-weight portfolio \eg test period 2019 and $T=60$. This might be attributed to D-Va not being able to capture any information on possible shocks to the price trends over the next $T$ days. However, for such cases, its performance often lies close to the equal-weight portfolio, showing that the regularization method is able to spread the risk out well when there are limited information. 

Finally, Table \ref{sharpe_final} compares the average $T$-day Sharpe ratios across the three benchmark models, after the graphical lasso regularization is applied. We see that our D-Va model is able to obtain the best results on Sharpe ratio, and results close to that of the equal-weight portfolio when it does not, which was observed previously. One important observation is that the Sharpe ratio results are not directly proportional to the prediction MSE results in Table \ref{results}. 
This is likely because the forming of portfolios take into account the volatility \textit{direction} of the predictions -- for example, for a true value of 0.0, a prediction of -0.01 and 0.01 would contribute the same impact to the MSE but would result in different trading decisions when forming a portfolio. However, our D-Va model was still able to achieve state-of-the-art results in both metrics. It is possible that dealing with the stochasticity of the target series allows the model to make "cleaner" predictions, which allows the MSE to accurately reflect its actual closeness to the target series and not its noise. 
\begin{table}[h]\small
\caption{Comparison of Sharpe ratios across benchmark multi-step prediction models, after the graphical lasso regularization is applied. The best results are boldfaced.}
\vspace{-5px}
\label{sharpe_final}
\begin{tabular}{llccccc}
\hline
\multicolumn{2}{c}{Model}                                            & \hspace*{0.5mm}ARIMA\hspace*{0.5mm}   & \hspace*{0.5mm}NBA\hspace*{0.5mm}    & Autoformer & \hspace*{0.5mm}Equal\hspace*{0.5mm}          & \hspace*{2mm}D-Va\hspace*{2mm}           \\ \hline
\multicolumn{1}{c|}{\multirow{4}{*}{2016}} & \multicolumn{1}{l|}{10} & 0.0953  & 0.0691  & 0.1114      & 0.1089          & \textbf{0.1174} \\
\multicolumn{1}{c|}{}                      & \multicolumn{1}{l|}{20} & 0.0815  & 0.1075  & 0.0921      & 0.1243          & \textbf{0.1246} \\
\multicolumn{1}{c|}{}                      & \multicolumn{1}{l|}{40} & 0.0683  & 0.0717  & 0.0438      & 0.0950          & \textbf{0.0953} \\
\multicolumn{1}{c|}{}                      & \multicolumn{1}{l|}{60} & 0.0544  & 0.0803  & 0.0235      & 0.0614          & \textbf{0.0929} \\ \hline
\multicolumn{1}{l|}{\multirow{4}{*}{2019}} & \multicolumn{1}{l|}{10} & 0.2736  & 0.1767  & 0.2350      & 0.2337          & \textbf{0.2767} \\
\multicolumn{1}{l|}{}                      & \multicolumn{1}{l|}{20} & 0.2455  & 0.1443  & 0.1672      & 0.1912          & \textbf{0.2467} \\
\multicolumn{1}{l|}{}                      & \multicolumn{1}{l|}{40} & 0.2152  & 0.1252  & 0.1031      & 0.1113          & \textbf{0.2152} \\
\multicolumn{1}{l|}{}                      & \multicolumn{1}{l|}{60} & 0.0732  & 0.0826  & 0.0804      & \textbf{0.0965} & 0.0914          \\ \hline
\multicolumn{1}{l|}{\multirow{4}{*}{2022}} & \multicolumn{1}{l|}{10} & 0.0173  & 0.0404  & 0.0544      & 0.0437          & \textbf{0.0645} \\
\multicolumn{1}{l|}{}                      & \multicolumn{1}{l|}{20} & 0.0408  & 0.0404  & 0.0141      & 0.0505          & \textbf{0.0540} \\
\multicolumn{1}{l|}{}                      & \multicolumn{1}{l|}{40} & 0.0287  & -0.0223 & -0.0016     & 0.0365          & \textbf{0.0527} \\
\multicolumn{1}{l|}{}                      & \multicolumn{1}{l|}{60} &  -0.0329  & -0.0179 & -0.0630     & -0.0230         & \textbf{0.0524} \\ \hline
\end{tabular}
\vspace{-20px}
\end{table}
\section{Conclusion and Future Work}
In this paper, we explained the importance of the multi-step regression stock price prediction task, which is often overlooked in current literature. For this task, we highlighted two challenges: the limitations of existing methods in handling the stochastic noise in stock price input data, and the additional problem of noise in the target sequence for the multi-step task. To tackle these challenges, we propose a deep-learning framework, D-Va, that integrates hierarchical VAE and diffusion probabilistic techniques to do multi-step predictions. We conducted extensive experiments on two benchmark datasets, one of which is collected by us by extending a popular stock prediction dataset \cite{xu2018stock}. We found that our D-Va model outperforms state-of-the-art methods in both its prediction accuracy and the standard deviation of the results. Furthermore, we also demonstrated the effectiveness of the model outputs in a practical investment setting. The portfolios formed from D-Va's predictions are also able to outperform those formed from the other benchmark prediction models and also the equal-weight portfolio, which is a known strong baseline in finance literature \cite{demiguel2009optimal}. 

The results of this work open up some possible future directions for research. On data augmentation, we have explored perturbing the data by the gradient of the loss function \cite{feng2018enhancing} and Gaussian diffusion noise in this work. Other possibilities include adding noise between the range of the high and low prices, which represent the maximum observed price movements for each time-step. A recent work \cite{ziyin2022theoretically} also proposed theoretically that augmenting stock data with noise of strength $\sqrt{r}$ is the most optimal to achieve the best Sharpe ratio. Additionally, there has also been numerous existing research on predicting stock movements with alternative data, such as text \cite{hu2018listening, feng2021time} or audio \cite{yang2022numhtml}. This could be incorporated into D-Va, which currently only uses the historical price data. However, instead of simply looking at prediction accuracy, one can also explore how the additional information sources help to reduce the \textit{uncertainty} of the predictions. This will be greatly helpful for the case of predicting highly stochastic data, such as stock prices. Finally, given the sequence prediction outputs from D-Va, other state-of-the-art portfolio techniques can also be explored, such as the Hierarchical Risk Parity (HRP) approach \cite{de2016building}, or the most recent Partially Egalitarian Portfolio Selection (PEPS) technique \cite{peng2022portfolio}, to evaluate the synergy of our model with different financial techniques. 
\vspace{-2px}
\section{Acknowledgement}
This research is supported by the Defence Science and Technology Agency, and the NExT Research Centre.

\bibliographystyle{ACM-Reference-Format}
\bibliography{reference}

\appendix

\end{document}